\documentclass[onecolumn,showpacs]{revtex4}
\usepackage{amssymb}
\usepackage{makeidx}
\usepackage{amsmath}
\usepackage{graphicx}
\usepackage{epstopdf}
\usepackage{dcolumn}
\usepackage{bm}

\input{tcilatex}
\begin{document}

\title{Second and third harmonics generation by coherent sub-THz radiation
at induced Lifshitz transitions in gapped bilayer graphene}
\author{A.G. Ghazaryan $^{a} $}
\email{amarkos@ysu.am}
\author{H.H. Matevosyan $^{b} $}
\author{Kh.V. Sedrakian$^{a} $}
\affiliation{$^{a} $ Centre of Strong Fields Physics, Yerevan State University, 1 A. Manukian,
Yerevan 0025, Armenia\\
$^{b} $ Institute Radiophysics and Electronics NAS RA, 1 Alikhanian brs., Ashtarak 0203, Armenia\\}
\date{\today }

\begin{abstract}
Using the microscopic nonlinear quantum theory of interaction of strong
coherent electromagnetic radiation with a gapped bilayer graphene is
developed for high harmonic generation at low-energy photon excitation-
induced Lifshitz transitions. The Liouville-von Neumann equation for the
density matrix is solved numerically at the nonadiabatic multiphoton
excitation regime. By numerical solutions, we examine the rates of the
second and third harmonics generation at the particle-hole annihilation in
induced Lifshitz transitions by the two linearly polarized coherent
electromagnetic waves propagating in opposite directions. The obtained
results show that the gapped bilayer graphene can serve as an effective
medium for generation of even and odd high harmonics in the sub-THz domain
of frequencies.
\end{abstract}

\maketitle



\section{Introduction}

Many quantum electrodynamic nonlinear phenomena induced by strong laser
radiation in condensed matter, specifically in graphene/nanostructures, have
significant contribution in low-energy physics and nano-opto-electronics and
have been systematically investigated mainly in case of monolayer graphene 
\cite{1,1bb,2} that has been conditioned by unique physical properties of
such two-dimensional (2D) nanosystem of atomic thickness \cite{1},  \cite{1bb}. On
the other hand, for induced electrodynamic phenomena in 2D atomic
systems-nanostructures a bilayer graphene ($AB$-stacked) is of great
interest too, since its electronic states are considerably richer than those
for monolayer graphene, and multiphoton resonant excitation with
high-harmonic generation (HHG) via nonlinear channels in bilayer graphene
were also considered \cite{hhg1,Abook,hhg2}. The pioneer studies of laser-induced
HHG process in recent decades were made generally in gaseous media, while it
is of interest the investigation of HHG and related processes in
low-dimensional nanostructures, such as graphene and its derivatives \cite%
{H2,Mer,Mer1,H3,H4,H6,H7,H8,H9,H99,H10,H11,H12,H12a,H13,H14,H15,H16,H17,H18,H19,H21}%
, bulk crystals \cite{s1,s2,s3,s5,s6,s7}, hexagonal boron nitride \cite{BN},
monolayer transition metal dichalcogenides \cite{TMD,TMD1,TMD11},
topological insulator \cite{TI}, \cite{TII}, in monolayers of black
phosphorus \cite{phosph}, buckled 2D hexagonal nanostructures \cite{Mer2019}%
, solids \cite{corcumsolid}, \cite{semimetal}, as well as in other 2D systems \cite%
{twodem1,twodem2,twodem3}. The 2D nanosystems enable to develop important
technological applications. The quantum cascade laser is one of such
examples \cite{QCL} using the physical phenomena in 2D systems such as the
quantum Hall effect \cite{Abook}. The nonlinear coherent response in $AB$%
-stacked bilayer graphene under the influence of intense electromagnetic
radiation leads to the modification of quasi-energy spectrum, the induction
of valley polarized currents \cite{26b}, \cite{27b} as well as the second and
third-order nonlinear-optical effects \cite{28b,29b,30b,31b}. Moreover, the
bilayer graphene system represents a unique system in which the topology of
the band structure can be externally influenced and chosen. The bilayer
graphene is a highly tunable material: not only one can tune the Fermi
energy using standard gates, as in single-layer graphene, but the band
structure can also be modified by external perturbations such as transverse
electric fields or strain \cite{1a,1ab,1aa,9,22b,24b,Liftrans}.
Particularly, due to the realization of a widely tunable electronic bandgap
in an electrically gated bilayer graphene \cite{9}, \cite{10,10a,16}, it is
of interest to consider HHG process in the strong wave-bilayer graphene
coupling regime with a bandgap induced by an external constant electric
field as a condensed matter material with nontrivial topology \cite{Mer2019}%
, in particular, with the bands acquiring Berry curvature \cite{Berry}.
Moreover, with the current technology \cite{16} one can make such large gaps
in $AB$-stacked bilayer graphene, which is sufficient to produce
field-effect transistors with a high on-off ratio not only at cryogenic
temperatures but at room-temperature \cite{Ber1}, \cite{transist}.

In addition, it is known the graphene-based low energy photon-counting
photodetector of different applications, in areas as diverse as medical and
space sciences or security applications. The tunable bandgap in bilayer
graphene may enable sensitive photon-counting photodetectors to operate with
a trade-off between resolution and operational temperatures, with resulting
operational benefits. Note that the large bandgap can also make possible
effective room temperature HHG \cite{H14} in bilayer graphene, which is
suppressed in intrinsic bilayer graphene \cite{H3}. Unfortunately, the
effective pump wave-induced Lifshitz transitions (with photon energy much
smaller than Lifshitz energy) in $AB$-stacked gapped bilayer graphene are
less investigated.

In the present paper, with the help of the numerical simulations in a
microscopic nonlinear quantum theory of interaction of $AB$-stacked gapped
bilayer graphene with the moderately strong laser radiation, we find out the
optimal values of the main parameters, in particular, for bandgap, pump wave
intensity, graphene temperature for practically significant case in
high-order harmonics coherent emission in the particle low-energy region of
the induced Lifshitz transitions (the fragmentation of the singly-connected
Fermi line into four separate pieces) \cite{22b}, \cite{Lifshitz,Lif1,Lif2}.
The Liouville-von Neumann equation is treated numerically for the generation
of the higher (here -second, third) harmonics in the multiphoton excitation
regime near the Dirac points of the Brillouin zone. We consider the harmonic
generation process in the nonadiabatic regime of interaction when the
Keldysh parameter is of the order of unity. The picture of the multiphoton
excitation of the Fermi-Dirac sea and the trigonal warping effect is also
revealed.  We examine the rates of the HHG at the particle-hole annihilation
in the strong effective field of two linearly polarized plane electromagnetic waves,
 propagating in opposite direction for
practically optimal parameters of the considering system. Due to the special
magnitudes of the bandgap, we can operate with the sample temperature. The
obtained results show that for specially chosen values of the corresponding
characteristic parameters of this process we can use gapped bilayer graphene
as a convenient nonlinear medium to generate the higher harmonics of the
pump wave with an effective yield in the sub-THz and THz domains of the
spectrum, at the graphene temperatures of higher than the cryogenic
temperatures.

The paper is organized as follows. In Sec. II the set of equations for a
single-particle density matrix is formulated and numerically solved in the
multiphoton interaction regime. In Sec. III, we consider the problem of
harmonics generation at the low-energy excitation of gapped bilayer
graphene. Finally, conclusions are given in Sec. IV.

\section{Basic theory}

At the higher harmonic emission process in gapped bilayer graphene during
the coherent electromagnetic radiation-induced Lifshitz transitions with the
energies much smaller than the Lifshitz energy $\mathcal{E}_{L}$ have some
peculiarities \cite{Lifshitz,Lif1,Lif2}, \cite{46bb,47bb,48bb,49bb}. The
Lifshitz transition someone is unique with thermoelectric properties of
bilayer graphene \cite{thermoelectric}. Two touching parabolas of the Fermi
surface are broken into four separate `pockets'. However, as opposed to bulk
graphite, the external perturbations like the strain \cite{19bb}, \cite{20bb} or
electric field \cite{21bb} can modify the topology of the electronic
dispersion and change the energy of the Lifshitz transition which connects
regions of different Fermi contour topologies \cite{22bb}. Moreover, due to
the two-dimensional nature of bilayer graphene, its chemical potential and
its topology can be tuned with electrostatic gates \cite{1}, simplifying
experimental studies of the Lifshitz transition. By the way, in unperturbed
bilayer graphene, it is achieved at low energies $\mathcal{E}_{L}\sim 1$ $%
\mathrm{meV}$. To induce the asymmetry, a chemical doping can be used \cite%
{1a} or external gates \cite{1ab} can be patterned. This induced asymmetry
opens a bandgap between the two layers of the bilayer graphene \cite{22b}, 
\cite{46bb,47bb,48bb,49bb}. As is shown in \cite{22bb}, particularly for an
induced asymmetry $U=100$ $\mathrm{meV,}$ the Lifshitz transition occurs at
the higher energy $\mathcal{E}_{L}\sim 1.6$ $\mathrm{meV}$. One can conclude
from the estimate given in \cite{22bb} that the experimental observation of
the Lifshitz transition should be facilitated by the layer-asymmetry induced
bandgap and that, the wider the gap, the more enhanced the visibility of
this effect.

At an intraband transitions the interaction of a particle with the wave at
the THz or sub-THz photon low-energies $\hbar \omega \ll \mathcal{E}_{L}$,
characterizes by the effective interaction parameter $\chi $ \cite{H3}:%
\begin{equation}
\chi =eE_{0}\mathrm{v}_{3}/(\hbar \omega ^{2}),  \label{00}
\end{equation}%
where $E_{0}$ is the wave strength, $\omega $ is the wave frequency, $e$ is
the electron charge. Due to the gap, the interband transitions are
characterized by so-called Keldysh \cite{Keld1,Keld2} parameter expressed in
the form: 
\begin{equation}
\gamma =\omega \sqrt{m_{\ast }U}/\left( eE_{0}\right) =\chi^{-1}\mathrm{%
v}_{3}\sqrt{m_{\ast }U}/\left( \hbar \omega \right) .  \label{01}
\end{equation}

Here $U$ is the bandgap energy, $\hbar $ is the Planck constant; $m_{\ast
}=\gamma _{1}/(2\mathrm{v}_{F}^{2})$ is the effective mass ($\mathrm{v}_{F}$
is the Fermi velocity in monolayer graphene); $\mathrm{v}_{3}=\sqrt{3}%
a\gamma _{3}/(2\hbar )\approx \mathrm{v}_{F}/8$ is the effective velocity
related to oblique interlayer hopping $\gamma _{3}=0.32$ $\mathrm{eV}$ ($%
a\approx 0.246$ $\mathrm{nm}$ is the distance between the nearest $A$
sites), $\gamma _{1}\simeq 0.39$ $\mathrm{eV.}$

For the gapped materials, the Keldysh parameter gives the character of the
ionization process which with the electron-hole pair creation is the first
step of HHG. In the limit of $\gamma \gg 1$, the multiphoton ionization
dominates in the ionization process. In the so-called nonadiabatic regime $%
\gamma \sim 1$, both multiphoton ionization and tunneling ionization can
take place. In the limit of $\gamma \ll 1$, the tunneling ionization
dominates. For the considered case, the ionization process reduces to the
transfer of the electron from the valence band into the conduction band that
is the creation of an electron-hole pair. Since the interband transitions
can be neglected when $\gamma \gg 1$, then the wavefield cannot provide
enough energy for the creation of an electron-hole pair, and the generation
of harmonics is suppressed. So that, in the nonadiabatic regime due to the
large ionization probabilities the intensity of harmonics can be
significantly enhanced compared with tunneling one \cite{H14}, \cite{H19}.
If $\gamma \sim 1$ or $\gamma \ll 1$, interband transitions take place. From
this point of view, condensed matter materials with bilayer graphene are
preferable due to the tunable bandgap with nontrivial topology.

In the present paper, we will consider the nonadiabatic regime for the
generation of HHG at $\chi \sim 1$ and $\gamma \sim 1$, when the multiphoton
effects become essential. Our consideration is mainly focused on the low
photon energies. Note that the average intensity of the wave expressed by $%
\chi $ can be given as\textrm{\ }%
\begin{equation*}
I_{\chi }=\chi ^{2}\times 6\times 10^{10}\mathrm{Wcm}^{-2}(\hbar \omega /%
\mathrm{eV})^{3},
\end{equation*}%
so the required intensity $I_{\chi }$\ for the nonlinear regime$\ $strongly
depends\textit{\ }on the photon energy. Particularly, for the photons with
the energies $0.4-0.9$ $\mathrm{meV,}$ the multiphoton interaction regime
can be achieved at the intensities $I_{\chi }=1-10^{2}\mathrm{\ Wcm}^{-2}$.
Note, that modern photonic-based THz and sub-THz (with energies $0.4-1.24$ $%
\mathrm{meV}$) sources include quantum cascade lasers and can achieve
admirable output powers, mainly at cryogenic temperatures, whereas used in
conjunction with nonlinear crystals can make microwatts of tunable
continuous wave THz at room temperature \cite{subTHz}. Unfortunately, all of
these sources offer impressive performance in their own ways, but none so
far are easily integrated into larger digital electronic systems, which is
arguably their biggest downfall for communication systems \cite{subTHz1}.

In the following we use the microscopic nonlinear quantum theory of
interaction of coherent electromagnetic radiation with gapped bilayer
graphene which was developed in \cite{H14}, \cite{H19}. To get the sub-THz
frequencies, let propose two linearly polarized plane electromagnetic waves
with carrier frequency $\omega $ and slowly varying amplitude of the
electric field $E_{a}f(t)$:%
\begin{equation}
E_{1}\left( t\right) =\widehat{\mathbf{e}}E_{a}f(t)\cos \left( \omega t+%
\mathbf{kr}\right) ,  \label{a}
\end{equation}%
\begin{equation}
E_{2}\left( t\right) =\widehat{\mathbf{e}}E_{a}f(t)\cos \left( \omega t-%
\mathbf{kr}\right) ,  \label{1a}
\end{equation}%
propagating in opposite directions in a vacuum. We assume that the pump wave
wavelength $\lambda \gg a$, where $a$ is the characteristic size of the
carbon system--the distance between the nearest $A$ sites (for the HHG this
condition is always satisfied). In considering case of a standing wave
formed by the laser beams (\ref{1a}), a significant input in the HHG process
is conditioned by the Dirac points situated near the stationary maxima of
the standing wave. For these points, the magnetic fields of the
counterpropagating waves cancel each other \cite{Abook}, \cite{physrevA2011}%
, \cite{plasma2011}. Since the HHG is essentially produced at the lengths $%
l\ll \lambda $ near the electric field maximums, we assume the effective
field in the plane of the graphene sheets ($XY$) to be: 
\begin{equation}
\mathbf{E}\left( t\right) =f\left( t\right) E_{0}\widehat{\mathbf{e}}\cos
\omega t,  \label{7}
\end{equation}%
where $E_{0}=2E_{a}$, and $\widehat{\mathbf{e}}$ is the unit polarization
vector. The pump wave slowly varying envelope is described by the function:%
\begin{equation}
f\left( t\right) =\left\{ 
\begin{array}{cc}
\sin ^{2}\left( \pi t/\mathcal{T}\right) , & 0\leq t\leq \mathcal{T}, \\ 
0, & t<0,t>\mathcal{T},%
\end{array}%
\right. ,  \label{8}
\end{equation}%
where $\mathcal{T}$ characterizes the pulse duration and is taken to be $%
\mathcal{T}=10\mathcal{T}_{0}$, $\phi $ is the pump wave polarization
parameter, $\mathcal{T}_{0}=2\pi /\omega $.

In $AB$-stacked gapped bilayer graphene, the low-energy excitations $%
\left\vert \mathcal{E}_{\sigma }\right\vert <\gamma _{1}\simeq 0.39$ $%
\mathrm{eV}$ in the vicinity of the Dirac points $K_{\zeta }$ can be
described by an effective single-particle Hamiltonian \cite{9,22b,24b}:

\begin{equation}
\widehat{H}_{\zeta }=\left( 
\begin{array}{cc}
\frac{U}{2} & g_{\zeta }^{\ast }\left( \mathbf{p}\right) \\ 
g_{\zeta }\left( \mathbf{p}\right) & -\frac{U}{2}%
\end{array}%
\right) ,  \label{1}
\end{equation}%
where 
\begin{equation}
g_{\zeta }\left( \mathbf{p}\right) =-\frac{1}{2m_{\ast }}\left( \zeta 
\widehat{p}_{x}+i\widehat{p}_{y}\right) ^{2}+\mathrm{v}_{3}\left( \zeta 
\widehat{p}_{x}-i\widehat{p}_{y}\right) ,  \label{2}
\end{equation}%
$\mathbf{\hat{p}}=\left\{ \widehat{p}_{x},\widehat{p}_{y}\right\} $\textbf{\ 
}is the electron momentum operator, $\zeta =\pm 1$ is the valley quantum
number. The diagonal elements in Eq. (\ref{1}) correspond to opened gap $U$.
The first term in Eq. (\ref{2}) gives a pair of parabolic bands $\mathcal{E}%
=\pm p^{2}/(2m_{\ast })$, and the second term connecting with $\gamma _{3}$
causes trigonal warping in the band dispersion. The spin and the valley
quantum numbers are conserved. There is no degeneracy upon the valley
quantum number $\zeta $ for the issue considered. However, since there are
no intervalley transitions, the valley index $\zeta $ can be considered as a
parameter.

The eigenstates of the effective Hamiltonian (\ref{1}) are the spinors,%
\begin{equation}
\Psi _{\sigma }(\mathbf{r})=\frac{1}{\sqrt{S}}|\sigma ,\mathbf{p}\rangle e^{%
\frac{i}{\hbar }\mathbf{pr}}  \label{3}
\end{equation}%
where%
\begin{equation}
|\sigma ,\mathbf{p}\rangle =\frac{1}{\sqrt{S}}\sqrt{\frac{\mathcal{E}%
_{\sigma }+\frac{U}{2}}{2\mathcal{E}_{\sigma }}}\left( 
\begin{array}{c}
1 \\ 
\frac{1}{\mathcal{E}_{\sigma }+\frac{U}{2}}\Upsilon \left( \mathbf{p}\right)%
\end{array}%
\right) ,  \label{4}
\end{equation}%
\begin{equation}
\Upsilon \left( \mathbf{p}\right) =-\frac{p^{2}}{2m}e^{i2\zeta \vartheta
}+\zeta \mathrm{v}_{3}pe^{-i\zeta \vartheta }.  \label{6}
\end{equation}%
$\vartheta =\arctan \left( p_{y}/p_{x}\right) $, $\sigma $ is the band
index: $\sigma =1$ and $\sigma =-1$ for conduction and valence bands,
respectively; and $S$ is the quantization area. The corresponding
eigenenergies are 
\begin{equation}
\mathcal{E}_{\sigma }\left( \mathbf{p}\right) =\sigma \sqrt{\frac{U^{2}}{4}%
+\left( \mathrm{v}_{3}p\right) ^{2}-\zeta \frac{\mathrm{v}_{3}p^{3}}{m}\cos
3\vartheta +\left( \frac{p^{2}}{2m}\right) ^{2}}.  \label{5}
\end{equation}%

\begin{figure}[tbp]
\includegraphics[width=.7\textwidth]{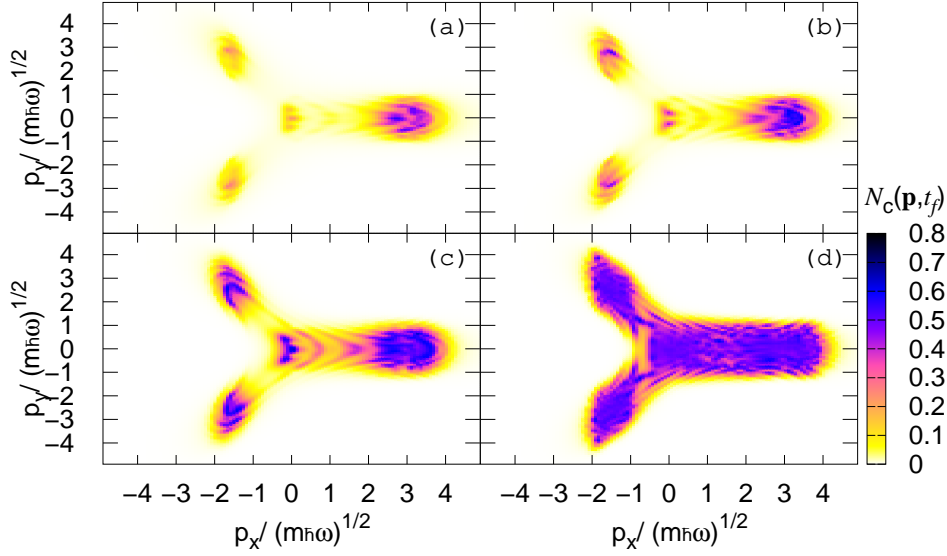}
\caption{(Color online) Particle distribution function $N_{c}(\mathbf{p}%
,t_{f})$\ (in arbitrary units) after the interaction at the instant $%
t_{f}=10T$, as a function of scaled dimensionless momentum components is
shown. The effective field is assumed to be linearly polarized along the $y$
axis. Multiphoton excitation with the trigonal warping effect for low-energy
photon excitation-induced Lifshitz transitions at the photon energy $\hbar 
\protect\omega =\mathcal{E}_{L}/1.1$\ $\simeq $\ $0.9$\ $\mathrm{meV}$, the
temperature $T/\hbar \protect\omega =0.4$ are demonstrated at dimensionless
intensity parameter $\protect\chi =1$ for valley $\protect\zeta =-1$.
(a)--(d) correspond to the gap energy $U=5$ $\mathrm{meV},$ $4$ $\mathrm{meV,%
}$ $3 $ $\mathrm{meV}$ and $0.0008$ $\mathrm{meV,}$ respectively.}
\label{111}
\end{figure}
Due to the second quantized technique, we can write the Fermi-Dirac field
operator in the form of an expansion in the free states, given in (\ref{3}),
that is, 
\begin{equation}
\widehat{\Psi }(\mathbf{r},t)=\sum\limits_{\mathbf{p,}\sigma }\widehat{a}_{%
\mathbf{p},\sigma }(t)\Psi _{\sigma }(\mathbf{r}),  \label{9}
\end{equation}%
where $\widehat{a}_{\mathbf{p},\sigma }(t)$ ($\widehat{a}_{\mathbf{p},\sigma
}^{+}(t)$) is the annihilation (creation) operator for an electron with
momentum $\mathbf{p}$ which satisfy the usual fermionic anticommutation
rules at equal times. The single-particle Hamiltonian in the presence of a
uniform time-dependent electric field $E(t)$ can be expressed in the form:%
\begin{equation}
\widehat{H}_{s}=\widehat{H}_{\zeta }+\left( 
\begin{array}{cc}
e\mathbf{rE}\left( t\right) & 0 \\ 
0 & e\mathbf{rE}\left( t\right)%
\end{array}%
\right) ,  \label{12}
\end{equation}%
where for the interaction Hamiltonian we have used a length gauge,
describing the interaction by the potential energy \cite{Corkum}, \cite{32b}. Taking
into account expansion (\ref{9}), the second quantized total Hamiltonian can
be expressed in the form: 
\begin{equation}
\widehat{H}=\sum\limits_{\sigma ,\mathbf{p}}\mathcal{E}_{\sigma }\left( 
\mathbf{p}\right) \widehat{a}_{\sigma \mathbf{p}}^{+}\widehat{a}_{\sigma 
\mathbf{p}}+\widehat{H}_{\mathrm{s}},  \label{13}
\end{equation}%
where the light--matter interaction part is given in terms of the
gauge-independent field $\mathbf{E}\left( t\right) $ as follow:

\begin{equation*}
\widehat{H}_{\mathrm{s}}=ie\sum\limits_{\mathbf{p,p}^{\prime },\sigma
}\delta _{\mathbf{p}^{\prime }\mathbf{p}}\partial _{\mathbf{p}^{\prime }}%
\mathbf{E}\left( t\right) \widehat{a}_{\mathbf{p},\sigma }^{\dagger }%
\widehat{a}_{\mathbf{p}^{\prime },\sigma ^{\prime }}\ 
\end{equation*}%
\begin{equation}
+\sum\limits_{\mathbf{p},\sigma }\mathbf{E}\left( t\right) \left( \mathbf{D}%
_{\mathrm{t}}\left( \sigma ,\mathbf{p}\right) \widehat{a}_{\mathbf{p},\sigma
}^{+}\widehat{a}_{\mathbf{p},-\sigma }+\mathbf{D}_{\mathrm{m}}\left( \sigma ,%
\mathbf{p}\right) \widehat{a}_{\mathbf{p},\sigma }^{+}\widehat{a}_{\mathbf{p}%
,\sigma }\right) ,  \label{14}
\end{equation}%
where 
\begin{equation}
\mathbf{D}_{\mathrm{t}}\left( \sigma ,\mathbf{p}\right) =\hbar e\langle
\sigma ,\mathbf{p}|i\partial _{\mathbf{p}}|-\sigma ,\mathbf{p}\rangle
\label{15}
\end{equation}%
is the transition dipole moment and 
\begin{equation}
\mathbf{D}_{\mathrm{m}}\left( \sigma ,\mathbf{p}\right) =\hbar e\langle
\sigma ,\mathbf{p}|i\partial _{\mathbf{p}}|\sigma ,\mathbf{p}\rangle
\label{16}
\end{equation}%
is the Berry connection or mean dipole moment, which are given in Appendix.

\begin{figure}[tbp]
\includegraphics[width=.7\textwidth]{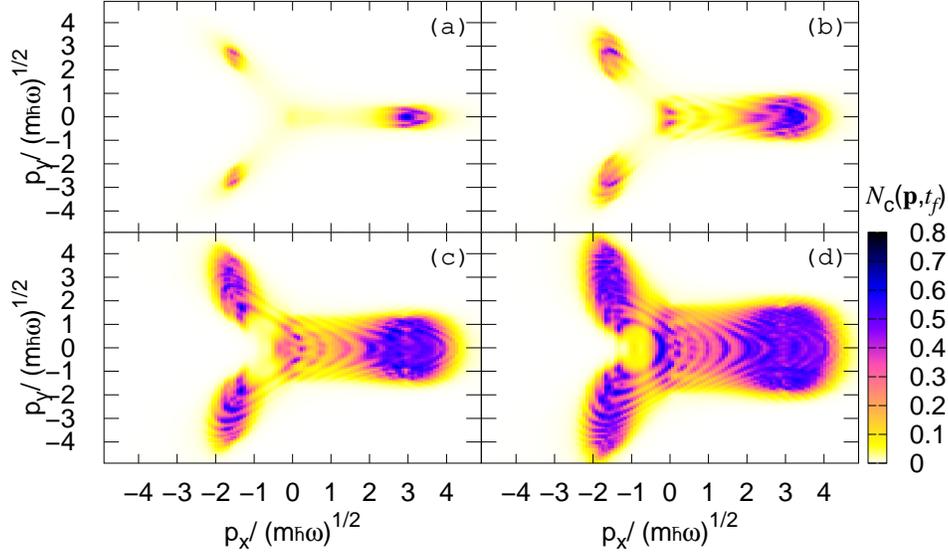}
\caption{ (Color online) Creation of a particle-hole pair in bilayer
graphene at multiphoton resonant excitation. Particle distribution function $%
N_{c}(\mathbf{p},t_{f})$ (in arbitrary units) after the interaction is
displayed at various intensities corresponding to : (a) $\protect\chi =0.5$,
(b) $\protect\chi =1$, (c) $\protect\chi =1.5$, and (d) $\protect\chi =2$.
The temperature is taken to be $T/\hbar \protect\omega =0.4$. The effective
field is assumed to be linearly polarized along the $y$ axis with the photon
energy $\hbar \protect\omega =\mathcal{E}_{L}/1.1$\ $\simeq $\ $0.9$\ $%
\mathrm{meV}$ and gap energy is $U=4$\ $\mathrm{meV}$. The results are for
the valley $\protect\zeta =-1$. }
\label{222}
\end{figure}

Multiphoton interaction of a bilayer graphene with a strong radiation field
will be described by the Liouville--von Neumann equation for a
single-particle density matrix (see Appendix equations (\ref{17}), (\ref%
{18})). As an initial state, we assume an ideal Fermi gas in equilibrium
with chemical potential to be zero. We will solve the set of Eqs. (\ref{19}%
), and followed from the last closed set of differential equations (\ref{20}%
), (\ref{21}) given in the Appendix, for the functions $N_{\mathrm{v}}(%
\mathbf{p},t)$, $N_{c}(\mathbf{p},t)$, $P(\mathbf{p},t)$, taking into
account the initial conditions:%
\begin{equation}
P(\mathbf{p},0)=0;\ N_{c}(\mathbf{p},0)=\frac{1}{1+e^{\mathcal{E}_{1}\left( 
\mathbf{p}\right) /T}};  \label{22a}
\end{equation}%
\begin{equation}
N_{\mathrm{v}}(\mathbf{p},0)=1-N_{c}(\mathbf{p},0).  \label{23}
\end{equation}%
Here $T$ is the temperature in energy units. Note that we will incorporate
relaxation processes into Liouville--von Neumann equation with inhomogeneous
phenomenological damping rate $\Gamma $, since homogeneous relaxation
processes are slow compared with inhomogeneous.

The set of equations (\ref{20}), (\ref{21}), and (\ref{22}) can not be
solved analytically. For the numerical solution we made a change of
variables and transform the equations with partial derivatives into ordinary
ones. The new variables are $t$ and $\widetilde{\mathbf{p}}=\mathbf{p}-%
\mathbf{p}_{E}$ $\left( t\right) $, where 
\begin{equation}
\mathbf{p}_{E}\left( t\right) =-e\int_{0}^{t}\mathbf{E}\left( t^{\prime
}\right) dt^{\prime }  \label{24}
\end{equation}%
is the classical momentum given by the wave field. After these
transformations, the integration of equations (\ref{20}), (\ref{21}), and (%
\ref{22}) is performed on a homogeneous grid of $10^{4}$ ($\widetilde{p}_{x},%
\widetilde{p}_{y}$)-points. For the maximal momentum we take $\widetilde{p}%
_{\max }/\sqrt{m\hbar \omega }=5$. The time integration is performed with
the standard fourth-order Runge-Kutta algorithm. For the relaxation rate we
take $\Gamma =0.5\mathcal{T}^{-1}$.

Photoexcitations of the Fermi-Dirac sea - induced Lifshitz transitions, are
presented in Figs. 1, 2. The effective wave is assumed to be linearly
polarized along the $y$ axis. Similar calculations for a wave linearly
polarized along the $x $\ axis show qualitatively the same picture. In Fig.1
density plot of the particle distribution function $N_{c}(\mathbf{p},t_{f})$%
\ is shown as a function of scaled dimensionless momentum components after
the interaction at the different energy gaps. The pump wave pulse duration
is $\mathcal{T}=10\mathcal{T}_{0}\approx 46$ $\mathrm{ps}$. It is clearly
seen as the trigonal warping effect describing the deviation of the excited
iso-energy contours from circles, which is smeared with the increase of the
gap magnitude. In all considering cases, the two touching parabolas are
transformed into the four separate \textquotedblleft
pockets\textquotedblright . Note that trigonal warping is crucial for
even-order nonlinearity. As is seen with the increasing of $U$ we approach
to perturbative regime $\gamma >1$ and only weak excitation of Fermi-Dirac
sea.
\begin{figure}[tbp]
\includegraphics[width=.7\textwidth]{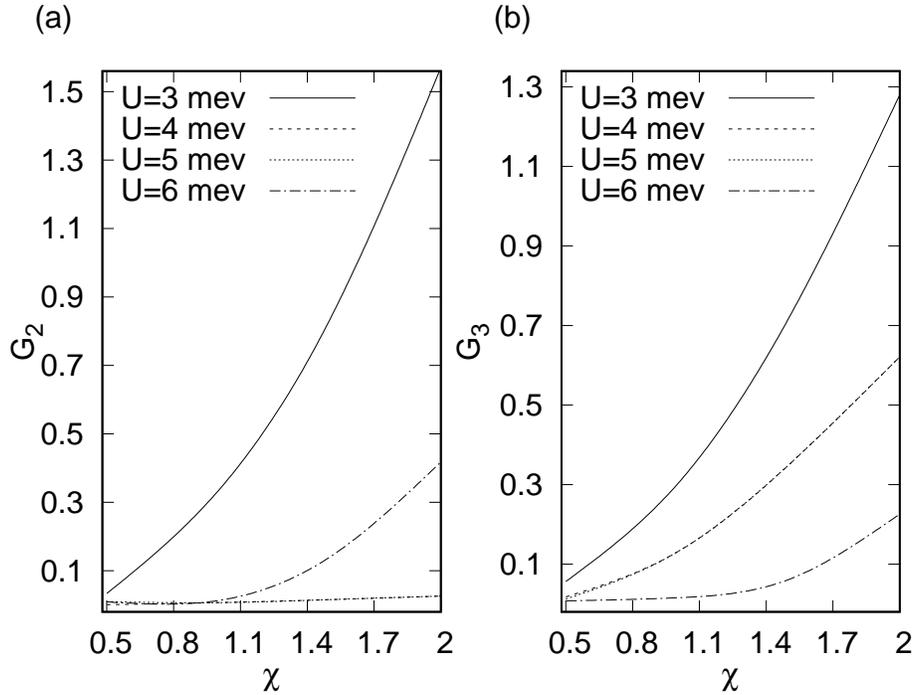}
\caption{The harmonic emission rate (in arbitrary units) of \ (a) second $%
G_{2}$ and (b) third $G_{3}$ order in bilayer graphene at Lifshitz
transition versus the intensity parameter $\protect\chi $ is shown for
various gap energies. The temperature is taken to be $T/\hbar \protect\omega %
=0.4$. The effective wave is assumed to be linearly polarized with the
frequency $\protect\omega =$\ $0.9$\ $\mathrm{meV/\hbar }$. }
\label{333}
\end{figure}

\begin{figure}[tbp]
\includegraphics[width=.7\textwidth]{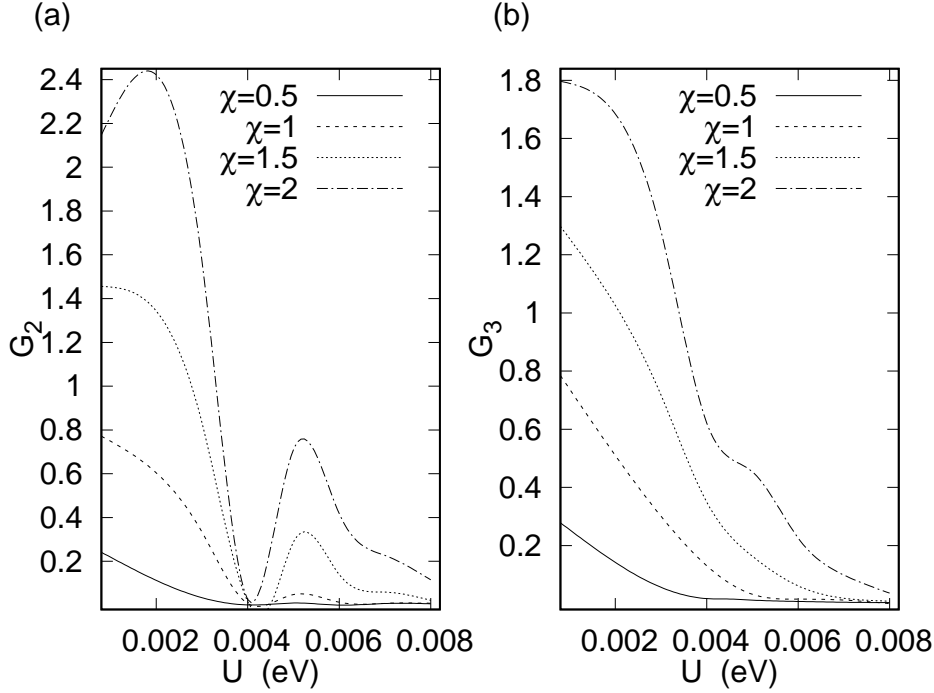}
\caption{(a) Second $G_{2}$ and (b) third $G_{3}$ harmonic emission rates
(in arbitrary units) versus the band gap $U$ for various intensities at the
same other parameters as in Fig. 3.}
\label{444}
\end{figure}

In Fig. 2, we show the photoexcitation depending on the pump wave intensity
at fixed sub-THz frequency. For the large values of $\chi $ when $\gamma
=1.1 $ we clearly see multiphoton excitations. With the increasing wave
intensity, the states with the absorption of more photons appear in the
Fermi-Dirac sea. At the parameters $\chi \succsim 1$ when $\gamma \simeq 1$,
the multiphoton excitation of the Fermi-Dirac sea takes place along the
trigonally warped isolines of the quasienergy spectrum modified by the wave
field. Thus, the multiphoton probabilities of particle-hole pair production
will have maximal values for the iso-energy contours defined by the resonant
condition $\mathcal{T}_{0}^{-1}\int\limits_{0}^{\mathcal{T}_{0}}2\mathcal{E}%
_{1}\left( \widetilde{\mathbf{p}}+\mathbf{p}_{E}\left( t\right) ,t\right)
dt=n\hbar \omega ,\ \ n=1,2,3$...These contours are also seen in Fig. 1. The
investigations of the temperature dependence of the excitation of the
Fermi-Dirac sea are cleared that for considering cases it exhibits a tenuous
dependence on the optimal temperatures: the excited isolines are slightly
smeared out with temperature increase. This effect is small since $U\gg $ $T$
and one can expect that harmonic spectra will be robust against temperature
change in contrast to the gap $U=0$ case where harmonics radiation is
suppressed with the increase of temperature in \cite{H14}, \cite{H19}. So,
the temperature dependence was missed.

In the following section, we will investigate the nonlinear response of the
bilayer graphene in the process of the second and third-order harmonics
generation under the influence of the laser field in the nonadiabatic regime 
$\gamma \simeq 1$ with the frequencies in sub-THz domain: $\omega =0.4\div
0.9$ $\mathrm{meV/}\hbar $.

\section{Harmonics generation at induced low-energy transitions in gapped
bilayer graphene}

In this section, we examine the nonlinear response of a bilayer graphene to
harmonic generation process considering the nonadiabatic regime of induced
Lifshitz transitions, when the Keldysh parameter is of the order of unity.
For the coherent part of the radiation spectrum one needs the mean value of
the current density operator,%
\begin{equation}
j_{\zeta }=-2e\left\langle \widehat{\Psi }(\mathbf{r},t)\left\vert \widehat{%
\mathbf{v}}_{\zeta }\right\vert \widehat{\Psi }(\mathbf{r},t)\right\rangle .
\label{50}
\end{equation}%
The velocity operator $\widehat{\mathbf{v}}_{\zeta }=\partial \widehat{H}%
/\partial \widehat{\mathbf{p}}$ is given in Appendix. Using the Eqs. (\ref%
{50})--(\ref{52}) and (\ref{17}), the expectation value of the current for
the valley $\zeta $ can be written in the form:%
\begin{equation*}
\mathbf{j}_{\zeta }\left( t\right) =-\frac{2e}{(2\pi \hbar )^{2}}\int d%
\mathbf{p}\left\{ \mathbf{V}_{1}\left( \mathbf{p}\right) \left( N_{c}(\mathbf{p}%
,t)-N_{\mathrm{v}}(\mathbf{p},t)\right) \right.
\end{equation*}%
\begin{equation}
\left. +2\hbar ^{-1}i\mathcal{E}_{1}\left( \mathbf{p}\right) \left[ \mathbf{D%
}_{\mathrm{t}}\left( \mathbf{p}\right) P^{\ast }(\mathbf{p},t)-\mathbf{D}_{%
\mathrm{t}}^{\ast }\left( \mathbf{p}\right) P(\mathbf{p},t)\right] \right\} ,
\label{53}
\end{equation}%
where $\mathbf{V}_{1}\left( \mathbf{p}\right) $ is the intraband velocity. In
Eq. (\ref{53}) the first term is the intraband current which conditioned by
intraband high harmonics and is generated as a result of the independent
motion of carriers in their respective bands. The second term in Eq. (\ref%
{53}) describes high harmonics which are generated as a result of the
recombination of accelerated electron-hole pairs. Since we study the
nonadiabatic regime, the contribution of both mechanisms is essential.

There is no degeneracy upon valley quantum number $\zeta $, so the total
current can be obtained by a summation over $\zeta $: 
\begin{equation}
j_{x}=j_{1,x}+j_{-1,x};  \label{55}
\end{equation}%
\begin{equation}
j_{y}=j_{1,y}+j_{-1,y}.  \label{56}
\end{equation}%
The current density components $j_{x,y}$ are defined as 
\begin{equation}
\frac{j_{x,y}}{j_{0}}=G_{x,y}\left( \omega t,\chi,\gamma ,\frac{%
\mathcal{E}_{L}}{\hbar \omega },\frac{T}{\hbar \omega },\frac{U}{\hbar
\omega }\right) .  \label{57}
\end{equation}%
Here 
\begin{equation}
j_{0}=\frac{e\omega }{\pi ^{2}}\sqrt{\frac{m\omega }{\hbar }},  \label{58}
\end{equation}%
and $G_{x}$ and $G_{y}$ are the dimensionless periodic (for monochromatic
wave) functions which parametrically depend on the interaction parameters $%
\chi $, $\gamma $, scaled Lifshitz energy, and temperature. Thus, using
solutions of \ Eqs. (\ref{20})-(\ref{22}), and making an integration in Eq. (%
\ref{53}), one can calculate the harmonic radiation spectra with the help of
a Fourier transform of the function $G_{x,y}(t)$. The emission rate of the $%
n $th harmonic is proportional to $n^{2}|j_{n}|^{2}$, where $|j_{n}|^{2}$ = $%
|j_{xn}|^{2}+|j_{yn}|^{2}$, with $j_{xn}$ and $j_{yn}$ being the $n$th
Fourier components of the field-induced total current. To find $j_{n}$, the
fast Fourier transform algorithm has been used. We use the normalized
current density (\ref{57}) for the plots.

For clarification of harmonics generation due to the multiphoton resonant
excitation and particle-hole annihilation, from the coherent superposition
states at $\gamma \simeq 1$ initially, we examine the emission rate of the
second and third harmonics. The emission rate versus the pump wave strength
defined by the parameter $\chi $ at the same wave frequency is demonstrated
in Fig. 3 for various gap energies. In Fig. 3, plots for $U=4$ $\mathrm{meV}$
and $U=5$ $\mathrm{meV}$ coincided. As is seen from this figure, for the
field intensities $\chi \succsim 1$ at the considering values of the gap
energy $U$ we have a strong deviation from power law for the emission rate
of the second or third harmonic (in accordance with the perturbation theory $%
\sim \chi ^{2}$ and $\chi ^{3}$, respectively). In Fig. 4, the second and
third harmonics emission rate, expectantly, is assumed as a function of the
energy gap at various intensities defined by the parameter $\chi $ at the
same wave frequency. As shown in Fig. 4, all plots have the minimum values
at $U=4$ $\mathrm{meV}$. As a result, we find the optimal parameters when
the harmonic emission rate is significant at the larger intensity for the
considered wave frequency. So, in accordance with the results of Figs. 3 and
4, the intense radiation of the second and third harmonics at the
pump-wave-induced particle or hole acceleration and annihilation in gapped
graphene can be obtained with the pump wave frequency in the sub-THz domain.
Then, as in case of the similar calculations for the intense pump wave or
large gap energy $U$ ($U\gg T$)\ has shown that the emission rate exhibits a
tenuous dependence on the temperature.

Figure 5 demonstrates the emission rate dependence on the pump frequency for
gapped bilayer graphene. We plot the second and third harmonic emission
rates for a gapped bilayer graphene versus the pump frequency for the gap $4$
$\mathrm{meV}$ and various intensity parameters. As was expected and shown
from this figure, the emission rate has different maxima at various photon
energies and intensities. For third harmonic generation, the maximal value
is reached at frequency $\omega =$\ $0.9$\ $\mathrm{meV/\hbar }$. Concerning
the generation of high harmonics up to the far-infrared range, note that
this has been demonstrated in the paper \cite{H19} where quantum cascade
lasers are readily available and can provide higher powers.

Finally, we investigate the physical conditions for considered case of HHG
in a bilayer graphene. Let us study the coherent interaction of bilayer
graphene with a pump wave in the ultrafast excitation regime, which is
correct only for the times $t<\tau _{\min }$, where $\tau _{\min }$\ is the
minimum of all relaxation times. For the excitations of energies $\mathcal{E}%
\ll \gamma _{1}=0.39$\ $\mathrm{eV}$, the dominant mechanism for relaxation
will be electron-phonon coupling via longitudinal acoustic phonons \cite%
{Hwang}, \cite{Viljas}. For the low-temperature limit $T\ll 2\left( c_{ph}/%
\mathrm{v}_{F}\right) \sqrt{\mathcal{E}\gamma _{1}}$, where $c_{ph}\simeq
2\times 10^{6}$\ $\mathrm{cm/s}$\ is the velocity of the longitudinal
acoustic phonon, the relaxation time for the energy level $\mathcal{E}$\ can
be estimated as \cite{Viljas}: 
 \begin{figure}[tbp]
\includegraphics[width=0.7\textwidth]{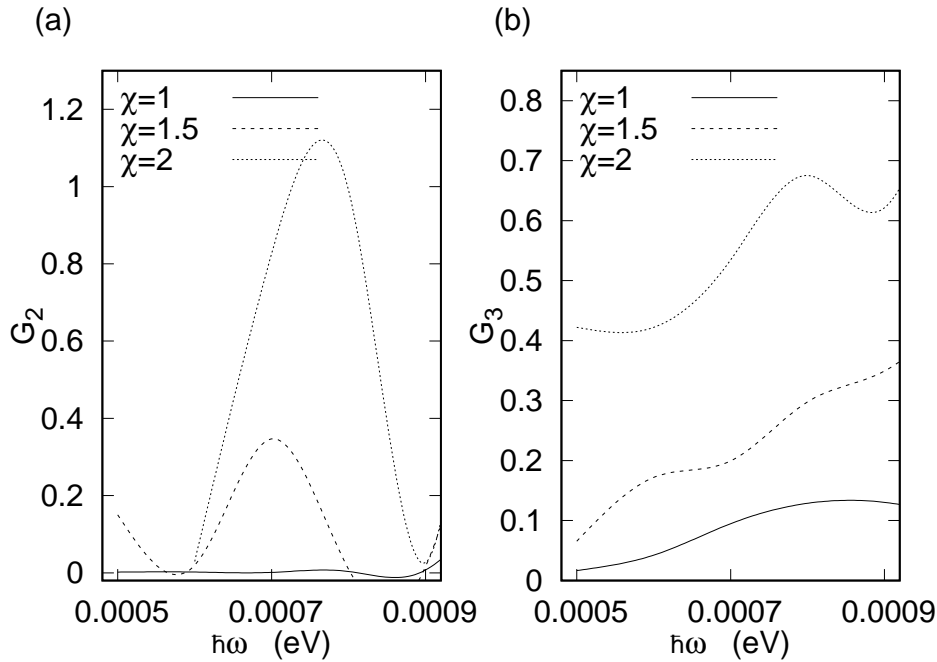}
\caption{(a) Second $G_{2}$ and (b) third $G_{3}$ harmonic emission
rate (in arbitrary units) for bilayer graphene versus $\hbar \protect\omega $
for various intensities. The temperature is taken to be $T/\hbar \protect%
\omega =0.4$, and the band gap $U=4$ $\mathrm{meV}$. The effective
field is assumed to be linearly polarized along the $y$ axis. }
\label{555}
\end{figure}
\begin{equation}
\tau \left( \mathcal{E}\right) \simeq \left( \frac{\pi D^{2}T^{2}}{8\rho
_{m}\hbar ^{3}c_{ph}^{3}\mathrm{v}_{F}}\sqrt{\frac{\gamma _{1}}{\mathcal{E}}}%
\right) ^{-1}.  \label{form1}
\end{equation}

Here $D\simeq 20$\ $\mathrm{eV}$\ is the electron-phonon coupling constant
and $\rho _{m}\simeq 15\times 10^{-8}$\ $\mathrm{g/cm}^{2}$\ is the mass
density of the bilayer graphene. For $\mathcal{E}\simeq 0.9$\ $\mathrm{meV}$
at the temperatures $T=0.4\hbar \omega $, from Eq. (\ref{form1}) we obtain $%
\tau \simeq 50$\ $\mathrm{ps}$. Thus, in this energy range one can
coherently manipulate with multiphoton transitions in bilayer graphene on
the time scales $t\precsim 50\mathrm{ps}$, not taking into account the
particle-particle collisions.

Note that the transition currents with comparing the intrinsic graphene $%
j_{0}$ \cite{Mer}, \cite{Mer1}, for a bilayer graphene is larger by a factor 
$\left( \gamma _{1}/2\hbar \omega \right) ^{1/2}$. Besides, the cutoff
harmonic is larger than in case of a monolayer graphene \cite{Mer}, which is
a result of strong nonlinearity caused by trigonal warping. Hence, for
considered setups $\hbar \omega \ll \gamma _{1}$ the harmonics' radiation
intensity is at least one order of magnitude larger than in the monolayer
graphene.

\section{Conclusion}

We have presented the microscopic theory of nonlinear interaction of the
gapped bilayer graphene with a strong coherent radiation field at low-energy
photon Lifshitz transitions. The closed set of differential equations for
the single-particle density matrix is solved numerically for a bilayer
graphene in the Dirac cone approximation. For the pump wave, the sub-THz
frequency range has been taken. Such waves can be produced by the two
linearly polarized plane electromagnetic waves of the same frequencies,
propagating in opposite. We have considered nonadiabatic wave-induced
Lifshitz transitions of Fermi-Dirac sea towards the HHG. It has been shown
that the role of the gap in the nonlinear optical response of bilayer
graphene is quite considerable. In particular, even-order nonlinear
processes are present in contrast to intrinsic bilayer graphene, the cutoff
of harmonics increases, and harmonic emission processes become robust
against the temperature increase. The obtained results show that the gapped
bilayer graphene can serve as an effective medium for the generation of even
and odd high harmonics in the THz and sub-THz domain of frequencies, which
is sufficient for a new high-speed wireless communication systems\
development \cite{subTHz}, \cite{subTHz1}. The obtained results certify that
the process of high-harmonic generation for sub-THz photons (wavelengths
from $0.3$ $\mathrm{mm}\ $to $1$ $\mathrm{mm}$) can already be observed for
intensities $I_{\chi }=1-10^{3}\mathrm{\ Wcm}^{-2}$ at the temperature of
the sample $T<\hbar \omega $. So, bilayer graphene is a very tunable
material. The features, such as the Lifshitz transition, can be largely
influenced by external parameters such as strain, magnetic fields, or even
displacement static electric fields, like to considering case. Bilayer
graphene represents therefore a unique system in which the topology of the
band structure can be externally influenced and chosen.

\section{Appendix}

Here we present the Liouville--von Neumann equation for a single-particle
density matrix%
\begin{equation}
\rho _{\alpha ,\beta }(\mathbf{p},t)=\langle \widehat{a}_{\mathbf{p},\beta
}^{+}\left( t\right) \widehat{a}_{\mathbf{p},\alpha }\left( t\right) \rangle
,  \label{17}
\end{equation}%
where $\widehat{a}_{\mathbf{p},\alpha }\left( t\right) $ obeys the
Heisenberg equation%
\begin{equation}
i\hbar \frac{\partial \widehat{a}_{\mathbf{p},\alpha }\left( t\right) }{%
\partial t}=\left[ \widehat{a}_{\mathbf{p},\alpha }\left( t\right) ,\widehat{%
H}\right] .  \label{18}
\end{equation}%
Note that due to the homogeneity of the problem we only need the $\mathbf{p}$%
-diagonal elements of the density matrix. Thus, taking into account Eqs. (%
\ref{13})-(\ref{18}), the evolutionary equation will be%

\begin{equation*}
i\hbar \frac{\partial \rho _{\alpha ,\beta }(\mathbf{p},t)}{\partial t}%
-i\hbar e\mathbf{E}\left( t\right) \frac{\partial \rho _{\alpha ,\beta }(%
\mathbf{p},t)}{\partial \mathbf{p}}=
\end{equation*}%
\begin{equation*}
\left( \mathcal{E}_{\alpha }\left( \mathbf{p}\right) -\mathcal{E}_{\beta
}\left( \mathbf{p}\right) -i\hbar \Gamma \left( 1-\delta _{\alpha \beta
}\right) \right) \rho _{\alpha ,\beta }(\mathbf{p},t)
\end{equation*}%
\begin{equation*}
+\mathbf{E}\left( t\right) \left( \mathbf{D}_{\mathrm{m}}\left( \alpha ,%
\mathbf{p}\right) -\mathbf{D}_{\mathrm{m}}\left( \beta ,\mathbf{p}\right)
\right) \rho _{\alpha ,\beta }(\mathbf{p},t)
\end{equation*}%
\begin{equation}
+\mathbf{E}\left( t\right) \left[ \mathbf{D}_{\mathrm{t}}\left( \alpha ,%
\mathbf{p}\right) \rho _{-\alpha ,\beta }(\mathbf{p},t)-\mathbf{D}_{\mathrm{t%
}}\left( -\beta ,\mathbf{p}\right) \rho _{\alpha ,-\beta }(\mathbf{p},t)%
\right] ,  \label{19}
\end{equation}%
where $\Gamma $ is the damping rate. In Eq. (\ref{19}) the diagonal elements
represent particle distribution functions for conduction $N_{c}(\mathbf{p}%
,t)=\rho _{1,1}(\mathbf{p},t)$ and valence $N_{\mathrm{v}}(\mathbf{p}%
,t)=\rho _{-1,-1}(\mathbf{p},t)$ bands, and the nondiagonal elements are
interband polarization $\rho _{1,-1}(\mathbf{p},t)=P(\mathbf{p},t)$ and its
complex conjugate $\rho _{-1,1}(\mathbf{p},t)=P^{\ast }(\mathbf{p},t)$.
Thus, we need to solve the closed set of differential equations for these
quantities:%

\begin{equation*}
i\hbar \frac{\partial N_{c}(\mathbf{p},t)}{\partial t}-i\hbar e\mathbf{E}%
\left( t\right) \frac{\partial N_{c}(\mathbf{p},t)}{\partial \mathbf{p}}=
\end{equation*}%
\begin{equation}
\mathbf{E}\left( t\right) \mathbf{D}_{\mathrm{t}}\left( \mathbf{p}\right)
P^{\ast }(\mathbf{p},t)-\mathbf{E}\left( t\right) \mathbf{D}_{\mathrm{t}%
}^{\ast }\left( \mathbf{p}\right) P(\mathbf{p},t),  \label{20}
\end{equation}%
\begin{equation*}
i\hbar \frac{\partial N_{\mathrm{v}}(\mathbf{p},t)}{\partial t}-i\hbar e%
\mathbf{E}\left( t\right) \frac{\partial N_{\mathrm{v}}(\mathbf{p},t)}{%
\partial \mathbf{p}}=
\end{equation*}%
\begin{equation}
-\mathbf{E}\left( t\right) \mathbf{D}_{\mathrm{t}}\left( \mathbf{p}\right)
P^{\ast }(\mathbf{p},t)+\mathbf{E}\left( t\right) \mathbf{D}_{\mathrm{t}%
}^{\ast }\left( \mathbf{p}\right) P(\mathbf{p},t),  \label{21}
\end{equation}%
\begin{equation*}
i\hbar \frac{\partial P(\mathbf{p},t)}{\partial t}-i\hbar e\mathbf{E}\left(
t\right) \frac{\partial P(\mathbf{p},t)}{\partial \mathbf{p}}=
\end{equation*}%
\begin{equation*}
\left[ 2\mathcal{E}_{1}\left( \mathbf{p}\right) +\mathbf{E}\left( t\right) 
\mathbf{D}_{\mathrm{m}}\left( \mathbf{p}\right) -i\hbar \Gamma \right] P(%
\mathbf{p},t)
\end{equation*}%
\begin{equation}
+\mathbf{E}\left( t\right) \mathbf{D}_{\mathrm{t}}\left( \mathbf{p}\right) %
\left[ N_{\mathrm{v}}(\mathbf{p},t)-N_{c}(\mathbf{p},t)\right] ,  \label{22}
\end{equation}%
The components of the transition dipole moments are calculated via Eq. (\ref%
{15}) by spinor wave functions (\ref{4}):%

\begin{equation*}
D_{\mathrm{t}x}\left( \mathbf{p}\right) =-\frac{e\hbar }{2\mathcal{E}%
_{1}\left( \mathbf{p}\right) \sqrt{\mathcal{E}_{1}^{2}\left( \mathbf{p}%
\right) -\frac{U^{2}}{4}}}
\end{equation*}%

\begin{equation*}
\mathbf{\times }\left( \left[ \left( \frac{p^{2}}{2m}-m\mathrm{v}%
_{3}^{2}\right) \frac{\zeta p_{y}}{m}+\frac{\mathrm{v}_{3}}{m}p_{x}p_{y}%
\right] \right.
\end{equation*}%
\begin{equation}
\left. -i\frac{U}{2\mathcal{E}_{1}}\left\{ \left( \frac{p^{2}}{2m}+m\mathrm{v%
}_{3}^{2}\right) \frac{p_{x}}{m}-\frac{3\zeta \mathrm{v}_{3}}{2m}\left(
p_{x}^{2}-p_{y}^{2}\right) \right\} \right) ,  \label{27}
\end{equation}%
\begin{equation*}
D_{\mathrm{t}y}\left( \mathbf{p}\right) =-\frac{e\hbar }{2\mathcal{E}%
_{1}\left( \mathbf{p}\right) \sqrt{\mathcal{E}_{1}^{2}\left( \mathbf{p}%
\right) -\frac{U^{2}}{4}}}
\end{equation*}%
\begin{equation*}
\times \left( \left[ \left( -\frac{p^{2}}{2m}+m\mathrm{v}_{3}^{2}\right) 
\frac{\zeta p_{x}}{m}+\frac{\mathrm{v}_{3}}{2m}\left(
p_{x}^{2}-p_{y}^{2}\right) \right] \right.
\end{equation*}%
\begin{equation}
\left. -i\frac{U}{2\mathcal{E}_{1}}\left\{ \left( \frac{p^{2}}{2m}+m\mathrm{v%
}_{3}^{2}\right) \frac{p_{y}}{m}+\frac{3\zeta \mathrm{v}_{3}}{m}%
p_{x}p_{y}\right\} \right) .  \label{28}
\end{equation}%
The total mean dipole moments are%

\begin{equation*}
D_{x\mathrm{m}}\left( \mathbf{p}\right) =D_{x\mathrm{m}}\left( 1,\mathbf{p}%
\right) -D_{x\mathrm{m}}\left( -1,\mathbf{p}\right) =-\frac{e\hbar U}{2%
\mathcal{E}_{1}\left( \mathbf{p}\right) \left( \mathcal{E}_{1}^{2}\left( 
\mathbf{p}\right) -\frac{U^{2}}{4}\right) }
\end{equation*}%
\begin{equation}
\mathbf{\times }\left[ \left( \frac{p^{2}}{2m}-m\mathrm{v}_{3}^{2}\right) 
\frac{\zeta p_{y}}{m}+\frac{\mathrm{v}_{3}}{m}p_{x}p_{y}\right] ,  \label{29}
\end{equation}%
\begin{equation*}
D_{y\mathrm{m}}\left( \mathbf{p}\right) =D_{y\mathrm{m}}\left( 1,\mathbf{p}%
\right) -D_{y\mathrm{m}}\left( -1,\mathbf{p}\right) =-\frac{e\hbar U}{2%
\mathcal{E}_{1}\left( \mathbf{p}\right) \left( \mathcal{E}_{1}^{2}\left( 
\mathbf{p}\right) -\frac{U^{2}}{4}\right) }
\end{equation*}%
\begin{equation}
\times \left[ \left( -\frac{p^{2}}{2m}+m\mathrm{v}_{3}^{2}\right) \frac{%
\zeta p_{x}}{m}+\frac{\mathrm{v}_{3}}{2m}\left( p_{x}^{2}-p_{y}^{2}\right) %
\right] .  \label{30}
\end{equation}%
Note, that the velocity operator is defined by the relation $\widehat{%
\mathbf{v}}_{\zeta }=\partial \widehat{H}/\partial \widehat{\mathbf{p}}$.
After the simple calculations for the effective $2\times 2$ Hamiltonian (\ref%
{1}), the velocity operator in components can be presented by the
expressions:%
\begin{equation}
\widehat{\mathrm{v}}_{\zeta x}=\zeta \left( 
\begin{array}{cc}
0 & -\frac{1}{m}\left( \zeta \widehat{p}_{x}-i\widehat{p}_{y}\right) +%
\mathrm{v}_{3} \\ 
-\frac{1}{m}\left( \zeta \widehat{p}_{x}+i\widehat{p}_{y}\right) +\mathrm{v}%
_{3} & 0%
\end{array}%
\right) ,  \label{51}
\end{equation}%
\begin{equation}
\widehat{\mathrm{v}}_{\zeta y}=i\left( 
\begin{array}{cc}
0 & \frac{1}{m}\left( \zeta \widehat{p}_{x}-i\widehat{p}_{y}\right) +\mathrm{%
v}_{3} \\ 
-\frac{1}{m}\left( \zeta \widehat{p}_{x}+i\widehat{p}_{y}\right) -\mathrm{v}%
_{3} & 0%
\end{array}%
\right) .  \label{52}
\end{equation}%
The intraband velocity $\mathbf{V}_{1}\left( \mathbf{p}\right) $ at HHG in
bilayer $AB$-stacked graphene is given by the formula: 
\begin{equation}
\mathbf{V}_{1}\left( \mathbf{p}\right) =\frac{\mathrm{v}_{3}\mathbf{p}-3\zeta 
\frac{\mathrm{v}_{3}p}{2m}\mathbf{p}\cos 3\vartheta +3\zeta \frac{\mathrm{v}%
_{3}p^{3}}{2m}\sin 3\vartheta \frac{\partial \vartheta }{\partial \mathbf{p}}%
+2\frac{\mathbf{p}^{3}}{\left( 2m\right) ^{2}}}{\mathcal{E}_{1}\left( 
\mathbf{p}\right) }.  \label{Vp}
\end{equation}

\begin{acknowledgments}
The authors are deeply grateful to prof. H. K. Avetissian for permanent discussions and valuable recommendations.
This work was supported by the RA MES Science Committee.
\end{acknowledgments}

\end{document}